\documentclass[twocolumn,backref, breaklinks, colorlinks]{aastex63}
\hypersetup{linkcolor=blue,citecolor=blue,filecolor=cyan,urlcolor=magenta}

\usepackage{tabularx}
\usepackage{url}
\usepackage{graphicx}
\usepackage[T1]{fontenc}
\usepackage{ae,aecompl}
\usepackage{booktabs}
\usepackage{natbib}
\usepackage{enumitem}
\usepackage{mathtools}
\usepackage{graphicx}
\usepackage{amssymb}
\usepackage{amsmath}
\usepackage{ulem}
\usepackage{epstopdf}
\usepackage{color}
\usepackage{graphicx}
\usepackage{xcolor}
\usepackage{needspace}

\usepackage{blindtext}
\usepackage{longtable}
\usepackage{tabu}
\usepackage{lineno}

\def\unit #1{\,{\rm #1}}

\newcommand\cmsqi{\rm \,\unit{cm^{-2}}}

\newcommand\cmcubei{\rm \,\unit{cm^{-3}}}

\newcommand\kev{\rm \,\unit{keV}}

\newcommand\funit{\rm \,erg\,cm^{-2}\,s^{-1}}
\newcommand\flu{\rm \,erg\,cm^{-2}}

\newcommand\lunit{\rm \,erg \,s^{-1}}

\newcommand\nh{\rm N_{H}}

\newcommand\mhz{\unit{MHz}}

\newcommand\pc{\unit{pc}}

\newcommand\mpc{\unit{Mpc}}
\newcommand\gpc{\unit{Gpc}}

\newcommand\chandra{{\it Chandra}}

\newcommand\frb{FRB~180301}
\newcommand\nicer{{\it NICER} {}}

\def\aap{A\&A} \def\apjl{ApJ}  
 \def\apjs{ApJS}

\shorttitle{FRB 180301}

\begin{document}
\tighten

\title{ Simultaneous view of the FRB~180301 with {\it FAST} and {\it NICER} during a bursting phase}

\author[0000-0003-2714-0487]{Sibasish Laha}

\affiliation{Center for Space Science and Technology, University of Maryland Baltimore County, 1000 Hilltop Circle, Baltimore, MD 21250, USA.}
\affiliation{Astrophysics Science Division, NASA Goddard Space Flight Center, Greenbelt, MD 20771, USA.}
\affiliation{Center for Research and Exploration in Space Science and Technology, NASA/GSFC, Greenbelt, Maryland 20771, USA}

\author[0000-0000-0000-0000]{George Younes} 
\affiliation{Astrophysics Science Division, NASA Goddard Space Flight Center, Greenbelt, MD 20771, USA.}

\author[0000-0003-2714-0487]{Zorawar Wadiasingh}
\affiliation{Department of Astronomy, University of Maryland, College Park, Maryland 20742, USA}
\affiliation{Astrophysics Science Division, NASA Goddard Space Flight Center, Greenbelt, MD 20771, USA.}
\affiliation{Center for Research and Exploration in Space Science and Technology, NASA/GSFC, Greenbelt, Maryland 20771, USA}

\author[0000-0000-0000-0000]{Bo-Jun Wang}
\affiliation{Kavli Institute for Astronomy and Astrophysics, Peking University, Beijing 100871, China}
\affiliation{Department of Astronomy, School of Physics, Peking University, Beijing 100871, China}
\affiliation{National Astronomical Observatories, Chinese Academy of Sciences, Beijing 100101, China}

\author[0000-0000-0000-0000]{Ke-Jia Lee}
\affiliation{Kavli Institute for Astronomy and Astrophysics, Peking University, Beijing 100871, China}
\affiliation{National Astronomical Observatories, Chinese Academy of Sciences, Beijing 100101, China}

\author[0000-0002-7465-0941]{Noel Klingler}
\affiliation{Center for Space Science and Technology, University of Maryland Baltimore County, 1000 Hilltop Circle, Baltimore, MD 21250, USA.}
\affiliation{Astrophysics Science Division, NASA Goddard Space Flight Center, Greenbelt, MD 20771, USA.}
\affiliation{Center for Research and Exploration in Space Science and Technology, NASA/GSFC, Greenbelt, Maryland 20771, USA}

\author[0000-0000-0000-0000]{Bing Zhang}
\affiliation{Nevada Center for Astrophysics, University of Nevada, Las Vegas, NV 89154, USA}
\affiliation{Department of Physics and Astronomy, University of Nevada Las Vegas, Las Vegas, NV 89154 USA.}

\author[0000-0000-0000-0000]{Heng Xu}
\affiliation{Kavli Institute for Astronomy and Astrophysics, Peking University, Beijing 100871, China}
\affiliation{Department of Astronomy, School of Physics, Peking University, Beijing 100871, China}
\affiliation{National Astronomical Observatories, Chinese Academy of Sciences, Beijing 100101, China}

\author[0000-0000-0000-0000]{Chun-Feng Zhang}
\affiliation{Kavli Institute for Astronomy and Astrophysics, Peking University, Beijing 100871, China}
\affiliation{Department of Astronomy, School of Physics, Peking University, Beijing 100871, China}
\affiliation{National Astronomical Observatories, Chinese Academy of Sciences, Beijing 100101, China}

\author[0000-0000-0000-0000]{Wei-Wei Zhu}
\affiliation{National Astronomical Observatories, Chinese Academy of Sciences, Beijing 100101, China}

\author[0000-0003-4790-2653]{Ritesh Ghosh}
\affiliation{Inter-University Centre for Astronomy and Astrophysics (IUCAA), Pune, 411007, India.}

\author[0000-0000-0000-0000]{Amy Lien}
\affiliation{University of Tampa, Department of Chemistry, Biochemistry, and Physics, 401 W. Kennedy Blvd, Tampa, FL 33606, USA}

\author[0000-0002-1869-7817]{Eleonora Troja}
\affiliation{University of Rome - Tor Vergata, via della Ricerca Scientifica 1, 00133, Rome, Italy}

\author[0000-0003-1673-970X]{S. Bradley Cenko}
\affiliation{Astrophysics Science Division, NASA Goddard Space Flight Center, Greenbelt, MD 20771, USA.}
\affiliation{Joint Space-Science Institute, University of Maryland, College Park, MD 20742, USA}

\author[0000-0000-0000-0000]{Samantha Oates}
\affiliation{School of Physics and Astronomy, University of Birmingham, Birmingham B15 2TT, UK}
\affiliation{Institute for Gravitational Wave Astronomy, University of Birmingham, Birmingham B15 2TT, UK}

\author[0000-0000-0000-0000]{Matt Nicholl}
\affiliation{School of Physics and Astronomy, University of Birmingham, Birmingham B15 2TT, UK}
\affiliation{Institute for Gravitational Wave Astronomy, University of Birmingham, Birmingham B15 2TT, UK}

\author[0000-0000-0000-0000]{Josefa Gonz\'alez ~Becerra } 
\affiliation{Instituto de Astrof\'isica de Canarias (IAC), E-38200 La Laguna, Tenerife, Spain}
\affiliation{Universidad de La Laguna (ULL), Departamento de Astrof\'isica, E-38206 La Laguna, Tenerife, Spain}

\author[0000-0000-0000-0000]{Eileen Meyer}
\affiliation{Department of Physics, University of Maryland, Baltimore County, 1000 Hilltop Circle, Baltimore, MD 21250, USA.}

\author[0000-0002-4299-2517]{Tyler Parsotan}
\affiliation{Center for Space Science and Technology, University of Maryland Baltimore County, 1000 Hilltop Circle, Baltimore, MD 21250, USA.}
\affiliation{Astrophysics Science Division, NASA Goddard Space Flight Center, Greenbelt, MD 20771, USA.}
\affiliation{Center for Research and Exploration in Space Science and Technology, NASA/GSFC, Greenbelt, Maryland 20771, USA}


\correspondingauthor{Sibasish Laha}
\email{sibasish.laha@nasa.gov,sib.laha@gmail.com}


\begin{abstract}
	
FRB180301 is one of the most actively repeating fast radio bursts (FRBs) which has shown polarization angle changes in its radio burst emission, an indication for their likely origin in the magnetosphere of a highly-magnetized neutron star. We carried out a multi-wavelength campaign with the FAST radio telescope and NICER X-ray observatory to investigate any possible X-ray emission temporally coincident with the bright radio bursts. The observations took place on 2021 March 4, 9 and 19. We detected five bright radio bursts with FAST, four of which were strictly simultaneous with the NICER observations. The peak flux-density of the radio bursts ranged between $28-105$ mJy, the burst fluence between $27-170$ mJy-ms, and the burst durations between $1.7-12.3$ ms. The radio bursts from FRB~180301 exhibited complex time domain structure, and subpulses were detected in individual bursts, with no significant circular polarisation. The linear degree of polarisation in L-band reduced significantly compared to the 2019 observations. We do not detect any X-ray emission in excess of the background during the 5~ms, 10~ms, 100~ms, 1~sec and 100~sec time intervals at/around the radio-burst barycenter-corrected arrival times, at a $>5\sigma$ confidence level. The $5\sigma$ upper limits on the X-ray a) persistent flux is  $<7.64\times 10^{-12}\funit$, equivalent to $L_{\rm X}<2.50 \times 10^{45}\lunit$ and b) 5 ms fluence is $<2\times 10^{-11}\flu$, at the radio burst regions. Using the $5$ ms X-ray fluence upper limit, we can estimate the radio efficiency $\eta_{R/X} \equiv L_{\rm Radio}/L_{\rm X-ray} \gtrsim 10^{-8}$. The derived upper limit on $\eta_{R/X}$ is consistent with both magnetospheric models and synchrotron maser models involving relativistic shocks.

\end{abstract}


\vspace{0.5cm}


\section{Introduction}\label{sec:intro}

Fast radio bursts (FRBs) are ms-duration radio pulses whose origin is still highly debated
\citep{Lorimer+07,Tendulkar+17,CHIME+19REPEATERS,petroff2019,zhang2020,2021Univ....7..453C}. Recently, large radio surveys have detected several new FRBs, some of which have shown repeating emission, implying an origin which does not involve one-time cataclysmic events, such as neutron star mergers \citep{CHIME+19REPEATERS}. These repeating FRBs are interesting because they may be scrutinized in different parts of the electromagnetic spectrum over long periods of time, to attempt to reveal the physical nature of the FRB engine. One such repeating FRB (but apparently not periodic in activity) in the CHIME/FRB catalog is FRB~20180301A (hereafter \frb{}), which was first detected by the Parkes 64-m radio telescope, and has a dispersion measure of $522 \cmcubei{} \pc{}$. \cite{bhandari2021} identified PSO~J093.2268+04.6703 as the putative host galaxy of \frb{}. The host of \frb{} is located at a redshift of $z=0.334$, implying a luminosity distance of $\sim 1.7\gpc{}$. The Five-hundred-meter Aperture Spherical radio Telescope (FAST), which is the largest single dish radio telescope with high sensitivity, observed this source in July, September, and October 2019 for a total of 12 hours \citep{luo2020}. The bursts detected from \frb{} had peak flux densities ranging from $5.3-94.1$ mJy. All the bursts exhibit a high degree of linear polarization, and no circular polarization was detected even for the highest signal to noise bursts. This property is similar to the other actively repeating FRB~121102, which exhibits $\sim 100\%$ linear polarization \citep{Michilli2018}. 

Most interestingly, a considerable amount of diversity in the polarization angle (PA) swings across the pulse profile were detected by FAST for \frb{}, which implies that the bursts are consistent with an origin from a neutron star magnetosphere and disfavors far-flung relativistic shocks \citep{luo2020}. The PA change from one burst to another in the same source indicates that the radiation is produced within the light cylinder of a strongly magnetized neutron star.  As the emitted radiation travels through the magnetosphere, the electric vector of the X-mode  wave adiabatically rotates and stays perpendicular to the local magnetic field direction (the O mode is what is approximately a normal mode of the plasma). The PA freezes at a radius where the plasma density becomes too small to be able to turn the electric vector. At the {\it freeze-out radius}, the electric field is perpendicular to the magnetic dipole moment of the neutron star projected in the plane of the sky, independent of the radiation mechanism or the orientation of the magnetic field in the emission region. The changes in the PAs from \frb{} should therefore, trace the rotational period of the underlying neutron star.

Magnetars have been historically strongly suspected as progenitors of FRBs \citep{2010vaoa.conf..129P,2013arXiv1307.4924P,2014ApJ...797...70K,2014MNRAS.442L...9L,2016ApJ...826..226K,2017ApJ...843L..26B,2017MNRAS.468.2726K,2018ApJ...868...31Y,Metzger+19,2019ApJ...879....4W}. The recent detection of FRB~200428 \citep{bochenek2020,CHIME-FRB200428} temporally coincident with a hard X-ray ($20-200\kev$) burst  \citep{mereghetti2020,li21} from magnetar SGR~1935+2154 (SGR~1935 hereafter) in April 2020 confirmed that at least some of the FRBs are produced by magnetar bursts. The radio to X-ray data have been interpreted within the magnetar framework in several competing models \citep[e.g.][]{lu2020,margalit2020,2020ApJ...903L..38W,2021ApJ...919...89Y,zhang2022}. On the other hand, the FRB luminosity detected from SGR~1935 is very low in comparison to its extragalactic counterparts, prompting yet another question of whether  Galactic and extragalactic FRBs indeed have the same origin. Nevertheless, the association of the FRBs with magnetars have prompted several dedicated searches of X-ray counterparts with current X-ray observatories. One of the important steps to test magnetar or shocked-outflow models is to estimate the ratio of energy emitted in the FRBs over that in other bands (such as in X-rays), measured by the efficiency factor $\eta_{R/X} \equiv E_{\rm radio}/E_{\rm Xray}$. In some magnetar models, this ratio is much less than unity, typically $\sim 10^{-3}$ to $10^{-7}$ \citep[e.g.][]{lu2020,margalit2020,2021ApJ...919...89Y}. Therefore strong limits on $\eta$ can challenge or vindicate several models.

Following the discovery of the PA changes in the source \frb{} \citep{luo2020},  we carried out a simultaneous radio-X-ray campaign in March 2021, with FAST and The Neutron Star Interior Composition Explorer Mission (NICER) telescopes to capture any X-ray emission temporally coincident with the radio bursts during the bursting phases of the FRB. There were five radio bursts from \frb{} during the multi-wavelength observational campaign and in this work we report a detailed analysis of the radio and X-ray observations. The paper is arranged as follows: Section \ref{Sec:obs} discusses the radio and X-ray observations and data analysis. Section \ref{Sec:results} lists the main results followed by discussion and summary in Section \ref{Sec:summary}, respectively.

\section{Observations and data reduction}\label{Sec:obs}

FAST \citep{NLC11} and NICER \citep{Gendreau2016} observed \frb \, on 2021 March 4, 9, and 19. See Tables \ref{Table:fast} and \ref{Table:nicer} for the FAST and NICER observation details, respectively. During these observations, four radio bursts were strictly contemporaneous. Below we describe the methods involved in data reprocessing and analysis of the radio and X-ray observations of \frb.

\subsection{FAST observations}
\label{sec:fast}

The radio observations were carried out using FAST, of which the effective collecting area is 196,000 m$^2$ \citep{NLC11}. We used the central beam of the 19-beam receiver to observe. In the frequency coverage of 1000-1500 MHz, the system temperature is 20 to 25~K \citep{NTL20}. Observed data were recorded using the digital backend based on the Re-configurable Open Architecture Computing Hardware-2 (\textsc{Roach2}) board \citep{HA16}, where the search data (i.e., the intensity or audio data) is formed via polyphase filterbanks and time integration on a \textsc{Xilinx} Virtex-6 family field-programmable gate array chip. The final temporal and frequency resolutions are 49.152 $\mu$s and 122.07 kHz respectively.

We searched for the FRB candidates offline with the recorded filterbank data. The two 20-MHz band edges, i.e., 1000-1020 MHz and 1480-1500 MHz were removed due to the sensitivity loss and rapid change of signal phase. Frequency channels, which were affected by satellite RFIs in 1200-1210 MHz and 1265-1280 MHz, were also removed. The software package \textsc{BEAR} (Burst Emission Automatic Roger) \citep{Men2019} is used to search for FRB candidates. Since FRB~180301 is a known repeater, we searched with a narrow DM range of 508 to 528~pc~cm$^{-3}$. We searched for pulses with the pulse width range of $0.2 - 30$ ms.
Candidates with S/N larger than 6 were recorded to evaluate the red noise effects as explained in \citet{CXM01}. Bursts with S/N larger than 7 will be visually inspected and reported in this paper. In this way, the chance of a burst being artifact is less than $3\times 10^{-6}$ assuming 10\% red noise power. Five bursts were detected in our observation with detailed information listed in Tab.~\ref{Table:fast}.

After detection, the DM is further refined using the phase coherence spectral techniques \citep{CAB19}, which optimize the burst sharpness instead of maximising the pulse S/N. 


We estimate the pulse flux assuming a 22 K system temperature, the major error of flux comes from the noise temperature variation, which is 20\% as measured in the FAST engineering phase. We calculate the mean flux using radiometer equation
\begin{equation}
S_{min}=\beta \frac{(\text{S/N})\cdot T_{sys}}{G\sqrt{N_{p} W_{eq} \Delta \nu}}
\end{equation}
where, $W$ is the pulse width, correction factor $\beta \approx~1$, and $N_{p}$ = 2 is the number of polarization channels. System temperature $T_{\mathrm{sys}}\approx~22~\mathrm{K}$ and gain $G=16 \mathrm{K~Jy^{-1}}$ for FAST. Neglecting the intrinsic bandwidth of FRB, we calculate, hereafter, the band-averaged flux and the bandwidth $\Delta\nu$ is fixed to be 400MHz. 

\begin{figure*}
    \centering
    \includegraphics[width=6 in]{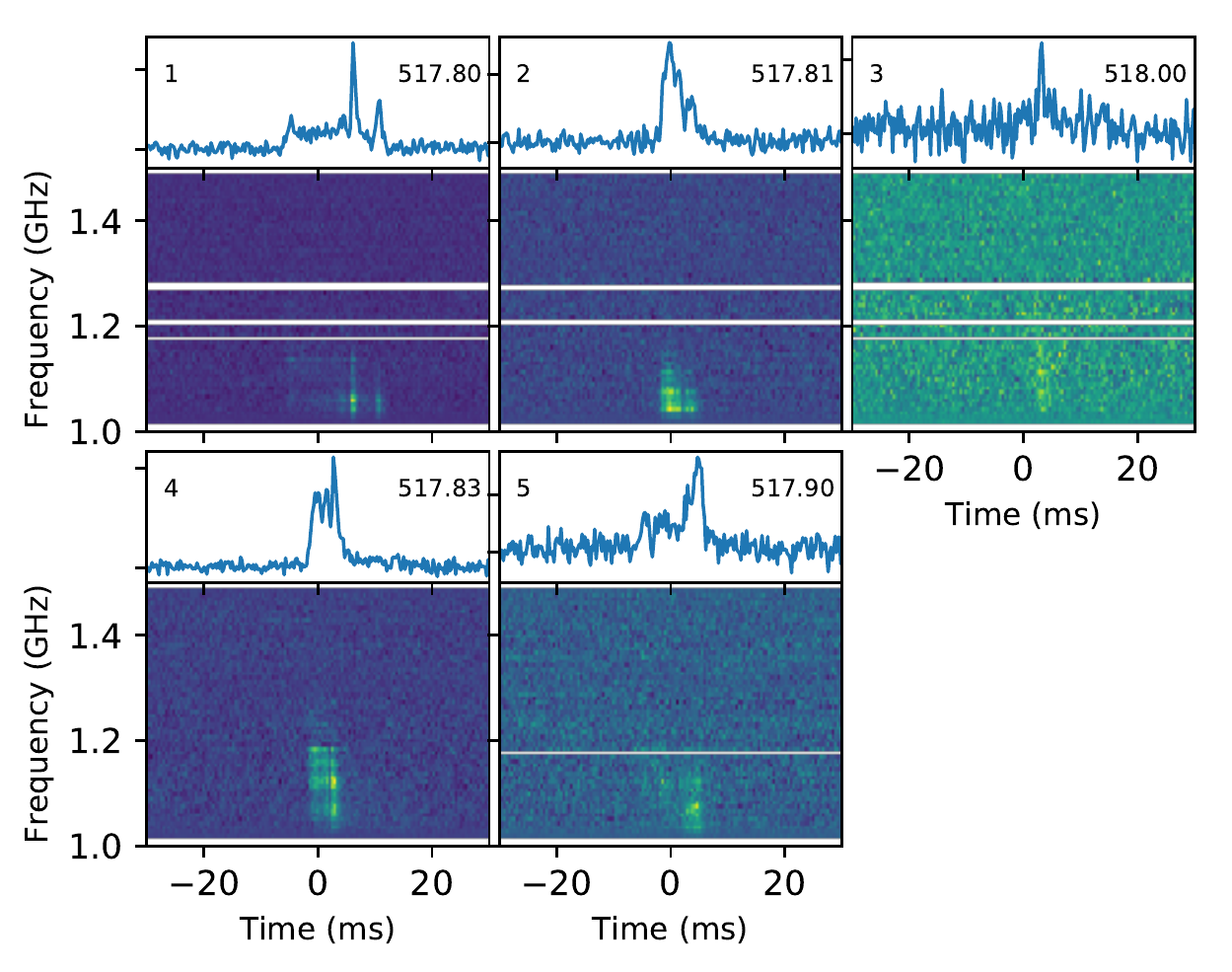}
    \caption{Pulse profile and dynamical spectra of FRB~180301 measured with FAST. The burst number (upper left) and DM (upper right) are given in each panel. White strips in the dynamical spectra indicate the RFI ``zapping". }
    \label{fig:fast}
\end{figure*}

We perform polarization calibration using software \textsc{Psrchive} with the single-axial model\citep{HVM04}, i.e. we neglected the leakage terms which is measured as low as -46 dB \citep{DBC17}. As will be discussed shortly after, the linear polarization is very weak compared to previous observations. We plot only the total intensity pulse profiles and de-dispersed dynamic spectra as in Fig.~\ref{fig:fast}.

The barycentric {\it{infinite-frequency equivalent}} time of arrivals are computed using \textsc{Tempo2} \citep{GEM06}, where R\"omer delay, relativity delay in Solar system, and dispersive time delay were corrected. In the process, we have adopted the position of RA $06^{\mathrm{h}}12^{\mathrm{m}}54.51^{\mathrm{s}}$, and DEC  $+04^\circ40'15.4''$ as measured with Karl G. Jansky Very Large Array (VLA) \citep{frbhost}. We understand that there may be an offset of approximately 2 mas between the \emph{The International Celestial Reference Frame} (ICRF) used by VLA and solar system dynamic coordinate used by Tempo2 \citep{WCH17},  the corresponding error in timing is 4.8 $\mu s$, which is negligibly small compare to the error of measured pulse width.

As in Fig~\ref{fig:fast}, the radio pulse profiles of FRB~180301 can be hardly described by Gaussian-like curves. We thus measure the pulse profile using the intensity weighted width (IWW), i.e. we treat the pulse profile as the temporal intensity distribution function, and calculate the standard deviation of time. A correction factor of $(8\ln2)^{1/2}$  is multiplied to the standard derivation when we report the pulse width. The factor is introduced such that the intensity weighted width will be the full width at half maximum (FWHM) for Gaussian profiles.

\subsection{NICER observations}
\label{sec:nicer}

NICER was launched in 2017 and is currently working as one of the payloads on the International Space Station (ISS). NICER consists of one instrument, the X-ray Timing Instrument (XTI), which operate in the soft X-ray band (0.2--12 keV).
The data files for the three NICER observations (PI: S.\ Laha, see Table \ref{Table:nicer}) were downloaded from HEASARC, and were reduced using the standard NICER procedure. 
The raw data were processed using the NICERDAS software package (version 2021-08-31\_V008c) in HEASOFT (v6.29c), using the latest caldb version.

We created cleaned event files by applying the standard calibration and filtering tool, {\tt nicerl2} to the unfiltered data using the default values, and then performed barycenter corrections using {\tt barycorr}. 
We restricted events to the 0.3--12 keV range.
To remove excess background noise from the time periods surrounding NICER's passages through the South Atlantic Anomaly (SAA), we binned the data into 16-second bins and filtered out intervals where the count rate exceeded 1.4 counts s$^{-1}$ in {\tt XSELECT}. This was done to obtain \nicer steady background rate. Although we note that two of the bursts (burst 1 and 3) happened in the wings when the spacecraft was coming-out of the SAA. 
We then used {\tt XSELECT} to extract light curves and spectra from the filtered data. 
We have used the latest response file, nixtiref20170601v002.rmf, for the spectral analysis. Note that we do not use the \nicer{} observation 1 for further analysis in this paper because it was not simultaneous with any radio burst from \frb{}, and also the duration was only for 700 sec, and hence the signal to noise was insufficient to carry out a detailed timing and spectral analysis.

We used Xselect to obtain the time averaged spectra for the two NICER observations. After plotting the source+background spectra, along with the modeled background spectra estimated using {\tt nicerbackgen} software, we do not find any excess emission above the background.



\section{Results}\label{Sec:results}

\subsection{The radio burst properties}

In the radio band, FRB~180301 exhibited complex time domain structure. Subpulses where detected in single bursts. As also detected in other repeating FRBs, subpulse frequency drifting is visible in the dynamic spectra of the burst No.4 and probably in the burst No.1 and 2. 

There is no significant circular polarization in the pulse we detected as in the previous observation \citep{luo2020}. Furthermore, we note that the linear degree of polarization in L-band reduced significantly in the March 2021 observations. The degree of linear polarization dropped to less than 10\%, while it was as high as 80\% in 2019.  The change of the observed polarization properties may be related to the propagation effects of the FRBs, and may probe the immediate environment around the FRB source \citep{xu2021,feng2022}. However, no significant change in the high energy emission properties, which depend on the intrinsic radiation mechanism, is expected. The detailed analysis of polarization properties is beyond the scope of the current paper and will be published in another work.

The mean and peak flux density of our observation to FRB~180301 range from 16 to 47 mJy and 28 to 105 mJy. The pulse width runs from 1.7 ms to 12 ms. Both of the value appears quite normal for FRB~180301. All five pulses are well above 7-$\sigma$ detection threshold. The minimal S/N of 11 indicates that the chance for any burst being spurious is less than $10^{-6}$ given the total observation of 13 ks even if 10\% correlated noise is included.


\begin{table*}
{\footnotesize
\caption{The list of simultaneous NICER observations of FRB~180301 in March 2021. }

\centering
	\label{Table:nicer}
	  \begin{tabular}{lcccccccc} \hline\hline 
		  NICER-obsid &Obs 		& Date			& Exposure  & FRB  \\
				    &	number	        & 	 		&(seconds)			&detected? \\ \hline 
				    
4533020101   &obs1   &2021-03-04              &   1998       & No  \\
4533020102      &obs2       &2021-03-09               &   4347      & Yes   \\
4533020103      &obs3       &2021-03-19            &   7151         &Yes  \\ \hline 
\end{tabular} 

}  
\end{table*}                                                            



\begin{table*}
{\footnotesize
\caption{The list of radio bursts from FRB~180301 detected by FAST during the joint FAST-NICER campaign in March 2021.}

\centering \label{Table:fast}
	  \begin{tabular}{lccccccc} \hline\hline 
Idx &	    Barycentric 		    & Day         & Peak 		&Mean    & Width    &S/N  \\
    &                               &             & flux density    &flux density &.         &   \\
N			&	TOA(MJD)                &             &mJy            &mJy        &ms         &	 \\ \hline
1          &59282.462719811028261    &09March2021          & 79        & 18   & 12.3$\pm$0.4 & 44       \\
2          &59282.519910811797640    &09March2021          & 44        & 16   & 4.1$\pm$0.2  & 21       \\
3          &59282.526543512274323    &09March2021       & 28        & 16   & 1.7$\pm$0.2  & 11       \\
4          &59292.407031067837291    &19March2021       & 105        & 47   & 4.10$\pm$0.06  & 69       \\
5          &59292.411839390156917    &19March2021     & 60        & 18   & 7.9$\pm$0.2  & 36       \\
\hline 
\end{tabular}               

Note: a) The major error of flux measurement comes from system temperature drift, which is approximately 20\%. b) the pulse width is defined as $(8\ln2)^{1/2}$ of the \emph{intensity weighted width}, which agrees with the definition of FWHM for Gaussian profiles.       

}                                                                                       
\end{table*}                                                                                     


\subsection{NICER Persistent Emission Flux Upper Limits} 

We do not detect any excess counts in the spectra above the NICER modeled background, which implies that the spectra is entirely dominated by background. Given this fact, we used the following steps to estimate the upper limit on the persistent flux for the source, for the given duration of the observations: Understanding that the observed spectra (for both the observations) are background dominated, we used  the following procedure to obtain the $5\sigma$ flux upper limit on a possible source detection. We note that the background count rate in the $0.2-10\kev$ in the region of the bursts 4 and 5 is $0.8$ counts/s, which implies a background flux of $2.2\times 10^{-12}\funit$ , using webpimms, with a power law slope of $\Gamma=2$, Galactic absorption column density of $\nh=2.83\times 10^{21}\cmsqi$ (Heasarc \citep{Kalberla2005}) and intrinsic absorption column of $\nh=10^{21}\cmsqi$ at $z=0.334$. A simple power law ($\Gamma=2$ frozen) fit to the observed time integrated spectrum of the source for observation 2 gives a $1\sigma$ error-on-background-flux $=1\times10^{-12}\funit$. Hence the net $5\sigma$ upper limit on the background, over which any signal registered can be confidently identified as a detection, can be estimated by: upper-limit=background-flux + 5$\times$error-on-background-flux+ $20\%$ of the background-flux. Note that we assumed $20\%$ of the background flux as systematic error, which is a conservative upper limit. Adopting the above prescription, the corresponding persistent flux upper limit is $<7.64\times 10^{-12}\funit$ over observations 2 and 3. The corresponding upper limit on the intrinsic persistent luminosity of the source turns out to be $L_{\rm 0.3-12\kev}< 2.50 \times 10^{45}\lunit$ for a luminosity distance of $1.7\gpc$.

\subsection{NICER Prompt Emission Flux Upper Limits}

 We searched carefully for any detectable X-ray counts around the radio burst arrival times, for the four strictly simultaneous bursts. \nicer{} is the only telescope with high effective area and with high temporal resolution in X-rays capable of capturing photon events even at sub-ms timescales. We binned the light curve to 1ms (typical FRB width) and searched for any photon counts in excess of $99.99\%$ confidence for any given bin, around the burst, for time intervals of 5~ms, 10~ms, 100~ms, 1~s and 100~sec. We did not find any excess counts above the background, at $>99.99\%$ confidence. In the next step we carried out a more rigorous and realistic simulation to estimate the $99.99\%$ confidence level of upper limit on detection at these five different time intervals around the burst times.

 We estimated the upper limit on the X-ray fluence of the four radio bursts coincident with \nicer{} observations using simulations which take into consideration the Poisson statistics and the average background count rate (that we have measured in each instance of the FRB). We note that the \nicer{} background count rate of $0.8$~counts/s were similar for bursts 4 and 5 and this was the time range which were not affected by the SAA flaring wings, as in the cases of bursts 1 and 3. Hence, we assumed the same X-ray background count rate for the bursts 1 and 3 which are contaminated by SAA entry-exit flaring wings. Therefore, we obtained only one set of upper limits for all the four bursts (albeit for different time resolutions). See Table \ref{Table:nicersims}. Below we illustrate the steps.

 Assuming a Poisson probability distribution we estimated the total number of source counts required in order to ``detect" a burst with $>99.99\%$ confidence, given a background, following the methods enumerated in \cite{gavriil2004,younes2020}. Since we are probing the five different timescales, we assumed the corresponding $\Delta T$ values of the bursts (i.e, $T_{90}$): 5ms, 10ms, 100ms, 1s and 100sec, and with a time resolution of investigation of $\sim 1/10$ of that of the $\Delta T$, in all the cases. This is to ensure that we are time-sampling the data adequately and not wash out the few counts in smaller time bins. We estimate the probability $P_{i}$ of the total counts in each time bin, $n_{i}$, to be a random fluctuation around the average value ($\lambda$), which is the ratio of the total counts within $\Delta T$ over $\Delta T$, as $P_{i}=(\lambda^{n_{i}}  exp(-\lambda))/n_{i}!$. The time bins satisfying the criterion $P_i<0.01/N$ are identified as a burst. The procedure is repeated until no more bins are identified in $\Delta T$. From the total source count rate $99.99\%$ upper limits obtained from the simulations, we converted it to flux using webpimms, assuming a spectral powerlaw slope of $\Gamma=2$.  The upper limits on the fluences are quoted in Table \ref{Table:nicersims}. For the case of $\Delta T=5$~ms we kept the time resolution = 2.5~ms. For the case of $\Delta T$=100~s we increased the background time to 100sec, and kept the time resolution 10 sec. In each case, we carried out 10,000 simulations.


\begin{table*}

\centering
	\caption{The X-ray flux and fluence upper-limits on different time scales of the four bursts that were simultaneous with NICER observations }\label{Table:nicersims}
	  \begin{tabular}{lllllllll} \hline\hline 
 		                &  5ms    	& 10ms & 100ms  &1 sec &  100 sec &			&     \\ \hline 

Counts  (N)                &   $11$    & $20$      & $29$      &$65$   & $700$  &    \\
Count rate (counts/sec) &   $2200$  &$2000$     & $290$     &$65$   & $7$  &      \\
Flux ($\funit{}$)       &   $4\times10^{-9}$  &$3.5\times10^{-9}$ & $4.9\times10^{-10}$ &$1.1\times10^{-10}$ & $1.2\times10^{-11}$  &      \\

Fluence upper limit ($\flu{}$) &   $2\times10^{-11}$  &$3.5\times10^{-11}$ & $4.9\times10^{-11}$ &$1.1\times10^{-10}$ & $1.2\times10^{-09}$  &      \\ \hline 
\end{tabular} 
$^{\rm a}$ For each case (5ms, 10ms, 100ms, 1000ms and 100sec) we have carried out multiple simulations step wise. Each simulation run had 10,000 simulations assuming a Poisson distribution of counts in each time bin. The input value of the simulation was the total number of counts (N), which we gradually increased in steps (for each simulation run) in order to achieve a detection probability of a possible burst at $99.99\%$ confidence for a background count rate of 0.8 counts/sec.  The quoted values of counts are those needed in that time interval in order for us to detect a burst at that confidence.                                                    
                                                                                       
\end{table*}

\subsection{The radio efficiency $\eta$}

Assuming a flat spectral index over a bandwidth of $\sim 200$ MHz, the FRB fluences in Table~\ref{Table:fast} are $\{6, 2, 0.7, 5, 4 \} \times 10^{-19}$ erg cm$^{-2}$. The corresponding dimensionless 10 ms transient fluence ratio lower limit is ${\cal F}_{\rm radio}/{\cal F}_{\rm X-ray} = \eta_{R/X} > \{1.7, - , 0.2, 1.4, 1.1 \} \times 10^{-8}$ adopting values in  Table~\ref{Table:nicersims}. The 10 ms transient limit represents a case similar to SGR 1935+2154, where offsets with the radio pulses (ToAs at infinite frequency equivalent) and features in the X-ray light curve were of order 7 ms \citep{mereghetti2020} and the width of the high-energy light curve features were $\sim 3$ ms. 

\begin{figure*}
    \centering
    \hbox{
    \includegraphics[width=9cm]{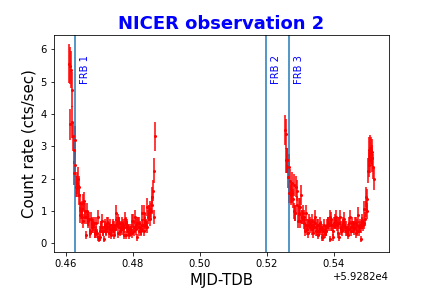}
    \includegraphics[width=9cm]{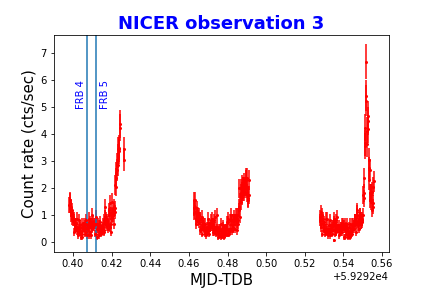}
    }
    \caption{The NICER light curve for observations 2 and 3.  The ``wings'' in the light curve are caused by periods of enhanced particle background as NICER enters/exits the SAA, and the gaps are due to periods of Earth occultation caused by the ISS's low Earth orbit.} 
    \label{fig:nicer1}
\end{figure*}

On the other hand, uncertainty in the DM of order $\Delta$DM $\sim 10$ pc cm$^{-3}$ may exist, which could impart temporal uncertainty of order $\sim 40$ ms for Figure~\ref{fig:nicer1}. However, we detect no unusual unmodeled fluctuations over background in any of the NICER snapshots consistent with activity from a cosmological source. 

The radio efficiency $\eta_{R/X}$ can in principle be used to differentiate the models involving magnetospheres or relativistic shocks, with the former models predicting a higher efficiency than the latter models \citep{zhang2020}. The derived upper limit, generally of the order of $\eta_{R/X} > 10^{-8}$, is however too loose to place a significant constraint, so that both magnetospheric models \citep[e..g.][]{2019ApJ...879....4W,2019MNRAS.488.5887S,2020arXiv200505093L,2020arXiv200102007L,lu2020,2020ApJ...903L..38W,2021arXiv210615030H,2021ApJ...919...89Y} and synchrotron maser models involving relativistic shocks \cite[e.g.,][]{2010vaoa.conf..129P,2014MNRAS.442L...9L,2017ApJ...843L..26B,Metzger+19,margalit2020} are allowed by the data.

\begin{figure*}[h!]
    \centering 
    {
    \includegraphics[width=18cm]{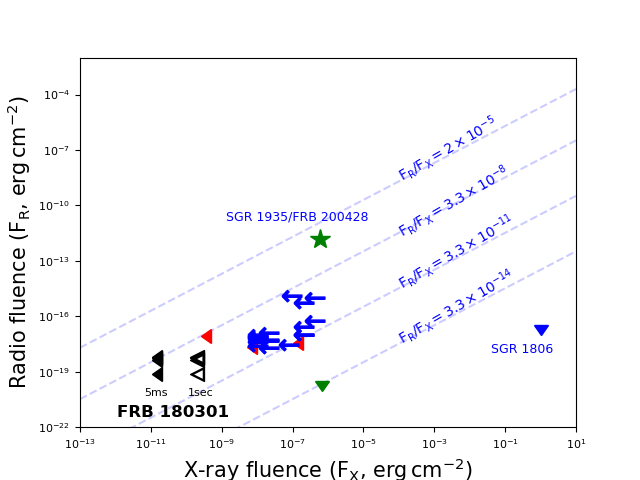}
    \caption{The X-ray and radio fluence limits of the FRBs and SGRs listed in Table \ref{Table:upper-limits}. The black solid triangles denote the X-ray upper limits of \frb{} from our work, using $5$ms integration time, while the hollow triangles refer to the limits when we used 1 sec integration time. The red triangles denote the X-ray upper limits for the three other repeating FRBs listed in Table \ref{Table:upper-limits}.  The blue arrows denote the X-ray fluence upper limits of the non-repeating FRBs listed in Table \ref{Table:upper-limits}. The green star is the contemporaneous detection in both X-rays and Radio of the SGR~1935/FRB~200428, while the green upper limit correspond to the radio fluence upper limits by FAST contemporaneous to 29 soft gamma bursts from SRG~1935 \citep{Lin2020}. The blue triangle in the extreme bottom right corner denotes the radio upper limit of the SGR~1806 during one of its bursting phases. References to all the studies are reported in the caption of Table \ref{Table:upper-limits}. }\label{fig:flu}
    }
    
\end{figure*}

\begin{figure*}[h!]
    \centering 
    {
    \includegraphics[width=18cm]{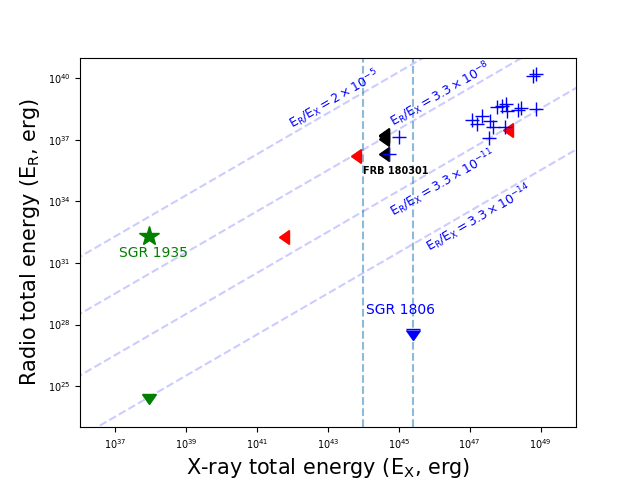}
    \caption{ The X-ray and radio energy limits measured for different FRBs and SGR as reported in Table \ref{Table:upper-limits}. The black triangles denote the X-ray upper limits of \frb{} from our work. The red triangles denote the X-ray upper limits for the repeating FRBs listed in Table \ref{Table:upper-limits}. The green star denotes the contemporaneous detection in either band for SGR~1935, while the green upper limit correspond to the radio fluence upper limits by FAST contemporaneous to 29 soft gamma bursts from SRG~1935 \citep{Lin2020}. The blue triangle in the bottom-right denotes the radio energy upper limit of the SGR~1806. Note that for all the repeating FRBs we have proper distance estimates. We have only upper limits on distance for the non-repeating FRBs, and hence we have plotted the corresponding values as a `+' sign, which denotes energy upper limit in both X (X-rays) and Y (Radio) axes.  The left dashed vertical line corresponds to a total energy of $10^{44}$ erg, and the right dashed vertical line refers to an energy of $2.45\times 10^{45}$ erg corresponding to the SGR~1806 giant flare and other similar local giant flares \citep{2021ApJ...907L..28B}. The references for the distances for all the sources are listed in the caption of Table \ref{Table:upper-limits}
    }\label{fig:energy}
    }
    
\end{figure*}

We searched the literature extensively to find cases where (1) an X-ray instrument was observing the source when the FRB was bursting, so as to obtain an X-ray upper limit contemporaneous with a radio burst, and (2) the FRB has a distance estimate or an upper limit from the dispersion measure. Table \ref{Table:upper-limits} shows the list of repeating and non-repeating FRBs selected from the literature using these criteria. We also include the two soft gamma repeaters (SGRs) for comparison. One is the classic case of SGR~1935, with simultaneous radio and X-ray detection \citep{mereghetti2020}. The other is the Galactic SGR~1806 which has shown giant flares, yet no contemporaneous signatures of FRB \citep{Tendulkar2016}, hence putting an strong upper limit on the radio fluence and energy. Figures \ref{fig:flu} and \ref{fig:energy}, we present the detections and upper limits of the various FRBs and SGR bursts with or without X-ray detections. In Fig \ref{fig:flu} the black triangles (filled and hollow) are the X-ray upper limits on the four radio bursts from \frb{} obtained in this work. The filled triangles are upper-limits corresponding to $5$ms integration time while the hollow triangles correspond to $1$sec integration time (also see Table \ref{Table:nicersims}). This is to have a fair comparison between the different X-ray instruments used to obtain the upper limits, which have different integration times. For example, \cite{scholz2020} obtained a $5\sigma$ prompt upper limit of $\sim 5\times 10^{-10}\flu$ for the FRB~180916, using \chandra{} which has an integration time of 3 seconds (marked as a red triangle in Fig \ref{fig:flu}). For most of the other cases in the literature, the integration time is of the order of a few ms. We plot two vertical lines in Fig \ref{fig:energy} which corresponds to the giant X-ray burst energy of the magnetars SGR~1806 ($\sim 2.5\times 10^{45}$ erg) and SGR~1900+14 ($\sim 10^{44}$ erg), to give a perspective of the energy involved. We find that the energy upper limits obtained from our work (black solid triangles) can rule out giant magnetar flares of the type detected in SGR~1806. However, for the non-repeating FRBs in the same Fig \ref{fig:energy} plotted as '+' (denoting upper limits in both X-rays and radio), the limits are not very constraining.

In Fig \ref{fig:flu}, the constant $\eta_{R/X}$ lines are also marked, and we can see that so far FRB 200428 from SGR 1935+2154 has the highest $\eta_{R/X}$, a value that could be interpreted within both the magnetospheric \citep{2019ApJ...876L..15W,lu2020,2021ApJ...919...89Y} and the external shock \citep{margalit2020} models. In order to make further progress to break the degeneracy between the models, simultaneous radio and X-ray observations of intrinsically bright FRBs at very small distances (in Milky Way or very nearby galaxies) are needed.

\begin{table*}
{\footnotesize
\caption{ The detection/upper-limits on the fluence/energy in radio and X-rays for different FRBs. }   \label{Table:upper-limits}  
\centering
	  \begin{tabular}{lcccccccccc} \hline\hline 
Source-Type &Source$^{\rm A}$  &	Radio fluence	& Radio frequency & X-ray fluence  & Distance$\rm ^B$   & Energy(Radio)     & Energy(X-ray) & \\
		&&($\flu{}^{\rm C}$)  &     &  $\flu$ & $\mpc$      &	$\rm erg$ & $\rm erg$ \\ \hline
	
Repeating   &   FRB~180301$^*$  & $6.5(-19)$ & 1.4GHz    & $<2(-11)$   &  $1700$ &   $1.68(37)$  &  $<5.20(44)$ \\
 FRB            & "             & $0.8(-19)$ & 1.4GHz    & $<2(-11)$   &  " &   $2.12(36)$  &   $<5.20(44)$\\
                &"              &$5.8(-19)$ & 1.4GHz    & $<2(-11)$   &  " &   $ 1.50(37)$  &   $<5.20(44)$ \\
                &"              &$4.3(-19)$ & 1.4GHz    & $<2(-11)$   &  " &   $ 1.10(37)$  &  $<5.20(44)$ \\
                &FRB~121102A$^{1,2}$ & $3.59(-18)$ & 1.4GHz   & $<2(-07)$  &  $949$ &   $2.92(37)$  &  $<1.62(48)$\\ 
                &FRB 180916$^{3}$     & $8.17(-18)$ & -    & $<5.00(-10)$   &  $140$ &   $1.53(36)$  &   $<8.82(43)$  \\ 
                &FRB~200120E$^4$&  $ 2.28(-18)$   &    -  &   $<1.00(-08)$ & $ 3$     & $1.85(32)$      & $<8.10(41)$    \\
                
                \hline		
Nonrepeating & FRB~010724$^1$& $4.49(-16)$ & 1.4GHz    & $<2(-07)$   & $<2000$ &   $<1.62(40)$  &  $<7.20(48)$  \\ 
FRB         &FRB~110220$^1$& $2.4(-17)$  &1.4GHz     & $<2(-07)$    &  $<1148$       &  $<2.84(38)$  &  $<2.37(48)$ \\ 
            &FRB~130729$^1$& $1.05(-17)$ &1.4GHz     & $<2(-08)$    &  $<2391$       &  $<5.40(38)$   & $<1.03(48)$     \\ 
            &FRB~010621$^1$& $8.7(-18)$  &1.4GHz     & $<2(-07)$    &  $<735$       & $<4.22(37)$      & $<9.72(47)$      \\ 
            &FRB~011025$^1$& $8.4(-18)$  &1.4GHz     &  $<2(-07)$   &  $<2029$       & $ <3.11(38)$     & $ <7.41(48)$   \\ 
            &FRB~131104$^1$& $8.1(-18)$  &1.4GHz     &  $<1(-08)$   &  $<1148$       & $ <9.6(37)$     & $ <1.18(47)$    \\ 
            &FRB~121002$^1$& $6.9(-18)$  &1.4GHz     &  $<1(-08)$   &  $<1558$        & $<1.5(38)$     & $<2.18(47)$  \\ 
            &FRB~090625$^1$& $6.6(-18)$  &1.4GHz     &  $<1(-08)$   &  $<2520$       & $<3.77(38)$     & $<5.71(47)$    \\ 
            &FRB~110703$^1$& $5.4(-18)$  &1.4GHz     &  $<1(-08)$   &  $<2980$       & $ <4.31(38)$      & $<7.99(47)$    \\
            &FRB~130626$^1$& $4.5(-18)$  &1.4GHz     &  $<2(-08)$   &  $<2520$       & $<2.57(38)$     & $<1.14(48)$    \\
            &FRB~140514$^1$& $3.9-18)$  &1.4GHz     &  $<2(-08)$    &  $<1148$       & $<4.62(37)$      & $<2.37(47)$   \\
            &FRB~130628$^1$& $3.59(-18)$  &1.4GHz     &  $<1(-08)$  &  $<1318$       & $<5.62(37)$     & $ <1.56(47)$    \\
            &FRB~110626$^1$& $2.1(-18)$  &1.4GHz     &  $<1(-08)$   &  $<2029$       & $<7.78(37)$      & $<3.70(47)$     \\
            &FRB~120127$^1$& $1.79(-18)$  &1.4GHz     &  $<2(-08)$  &  $<1609$       & $ < 4.19(37)$      & $<4.65(47)$    \\ 
            
             &FRB~180924B$^5$&  $4.79(-17)$         & -  & $<4.00(-07)$   &$ 896$     & $3.46(38)$      & $ <2.89(48)$ \\ 
             
             
             
             
             &FRB~190714A$^6$&  $2.4(-18)$  & - &    $<7.38(-08)$       &  $731$    & $1.15(37)$      & $<3.54(47)$   \\
             
             
             &FRB~171020A$^6$& $ 1.12(-15)$ &  -  & $<9.00(-08)$           & $36$     & $1.32(37)$      & $<1.04(45)$   \\
             &FRB~190523A$^7$& $8.39(-16)$ &  -  &   $<4.00(-07)$        & $1298$     & $1.27(40)$      & $<6.06(48)$  \\


            \hline 

SGR             &SGR~1086$^1$         &$<3.3(-17)$   &1.4GHz & $1.0(00)$ &  $ 0.014$        & $<5.82(27))$     & $2.46(45)$    \\
                &SGR~1935$^8$         & $1.44(-12)$  &  -    & $ 6.1(-07) $ &  $0.004 $       & $ 2.07(32)$     & $ 8.78(37)$   \\
                
                & " \hspace{0.5cm}$^9$         & $ <3(-20)$  &  -    & $ 6.8(-07) $ &  $0.004 $       & $ <4.3(24)$     & $ 9.00(37)$   \\


\hline \\
\end{tabular}               
$^*$ This work.\\
$\rm ^A$ References for Radio and X-ray fluences. Note that we quote the X-ray fluence upper limits from those studies where they have been derived contemporaneous with a radio burst from the respective FRB, except for FRB~200120E from M81. $^1$ \cite{Tendulkar2016}, 
$^2$ \cite{bhandari2021}, 
$^3$\cite{scholz2020}
$^4$\cite{majid2021,mereghetti2021}  
$^5$\cite{gourdji2020}
$^6$\cite{anumarlapudi2020}
$^7$\cite{Prochaska231},  
$^8$\cite{mereghetti2020} 
$^9$\cite{Lin2020}   \\ 

\vspace{0.5cm}

 $\rm ^B$ References for Distances: FRB~180301: \cite{bhandari2021} , FRB~121102: \cite{Tendulkar+17} , FRB~20201124A: \cite{fong2021}, FRB 180916: \cite{Marcote2020}. FRB~010724: \cite{guidorzi2019}, FRB~110220: \cite{petroff2015}  , FRB~130729: \cite{champion2016} , FRB~010621: \cite{keane2012}, FRB~011025: \cite{burke2014}, FRB~131104: \cite{sakamoto2021},  FRB~121002: \cite{champion2016}, FRB~090625: \cite{champion2016}, FRB~110703: \cite{Thornton+13}, FRB~130626: \cite{champion2016}, FRB~140514: \cite{petroff2015}, FRB~130628: \cite{champion2016}, FRB~110626: \cite{Thornton+13}, FRB~120127: \cite{Thornton+13}, FRBs 180924B, 190608B, 190714A, 171020A, 190523A, 18112: \cite{bhandari2021}.   \\ 
\vspace{0.5cm}

$\rm ^C$ Conversion of $\rm Jy-ms$ fluence to $\flu$: We know $\rm 1 Jy=10^{-23} erg/s/cm^{2}/Hz$, and we assume a bandwidth of $\sim \rm 300~MHz$. Hence for example, $(10^{-3}\times 18 \rm Jy) \times (12\times 10^{-3} \rm sec)\times (10^{-23} \rm erg/s/cm^{-2}/Hz/Jy) \times (3\times 10^8\rm Hz) = 6.5\times 10^{-19} \flu$ . For {\it Parkes telescope} we have used a band width of $288\mhz{}$ \citep{Tendulkar2016}, and a flat spectral index, to obtain the fluence in radio band at 1.4GHz. For the cases where we do not know the band width, we assume 300 MHz, and a flat spectral slope.\\

}                                                                                       
\end{table*}                                                                                     


\section{Summary and discussion}\label{Sec:summary}

In this work, we have reported detection of five bursts from \frb with FAST in L-band, four of which were contemporaneous with NICER monitoring of the source. The five bursts all exhibit no circular polarization, albeit with lower linear polarization than the previously reported epoch in 2019. No unusual X-ray emission over background is discernible for the four bursts which were strictly simultaneous (after accounting for DM) with NICER observations. Nor is any emission over background detected for possible delays associated with DM uncertainty or intrinsic temporal offsets, within the duration commensurate the NICER snapshots ($\sim 1000$ seconds). The corresponding radio-to-X-ray 5 ms fluence ratio is $\gtrsim 10^{-8}$ for the four strictly simultaneous radio bursts. An absorbed $5\sigma$ $0.3-12 \kev{}$ persistent X-ray flux limit of $L_{\rm 0.3-12\kev} < 6.0 \times 10^{45}\lunit$ is also found.  



A fluence of few$\times 10^{-19}$ erg cm$^{-2}$ yields a characteristic isotropic-equivalent radio energy of $10^{37}- 10^{38}$ erg, an energy scale that is typical for FRBs (repeaters and apparent nonrepeaters) with accurate localizations such as  FRB 121102, FRB 180924, FRB 181112 and FRB 20201124A \citep{Tendulkar+17,Bannistereaaw5903,Prochaska231, 2021ATel14516....1K,2021arXiv210909254L,piro2021}. This is consistent with models which predict narrow universal luminosity function \citep[e.g.,][]{Wadiasingh2020,2020MNRAS.tmp.1934B} for FRBs.

Our persistent and transient X-ray upper limits constrain any FRB progenitor to prompt radio efficiency $\eta_{R/X} \gtrsim 10^{-8}$. This is generally unconstraining for a large class of models involving stellar-mass compact objects and repetition. In the context of magnetar models, the persistent luminosity limit of $L_{\rm 2-10\kev}< 8 \times 10^{44}$ erg cm$^{-1}$ disfavors a scenario where separate magnetar giant flares are associated with each radio burst, if they produce bright quasi-thermal pulsations as known to follow the hard spike of giant flares in nearby magnetars \citep{1979Natur.282..587M,1980ApJ...237L...1C,1999Natur.397...41H,2005Natur.434.1107P,Hurley2005,2021ApJ...907L..28B}. The derived upper limit on $\eta_{R/X}$ is consistent with both magnetospheric models and synchrotron maser models involving relativistic shocks.


 \clearpage               
\acknowledgements
The material is based upon work supported by NASA under award number 80GSFC21M0002. MN is supported by the European Research Council (ERC) under the European Union’s Horizon 2020 research and innovation programme (grant agreement No. 948381) and by a Fellowship from the Alan Turing Institute.

\bibliographystyle{aasjournal}
\bibliography{mybib}
\end{document}